# Improved training in paraffin-wax impregnated Nb$_3$Sn Rutherford cables demonstrated in BOX samples

Michael Daly[1], Bernard Auchmann[1], André Brem[1], Christoph Hug[1], Serguei Sidorov[1],

Simon Otten[2], Marc Dhallé[2], Zichuan Guo[2], Anna Kario[2], Herman ten Kate[2]

[1] Paul Scherrer Institute, Villigen, Switzerland
[2] University of Twente, Faculty of Science and Technology, Enschede, The Netherlands

E-mail: michael.daly@psi.ch

**Abstract**

Resin-impregnated high-field Nb$_3$Sn type of accelerator magnets are known to require extensive training campaigns and even may exhibit performance-limiting defects after thermal or electromagnetic cycling. In order to efficiently explore technological solutions for this behaviour and assess a wide variety of impregnation material combinations and surface treatments, the BOnding eXperiment (BOX) sample was developed. BOX provides a short-sample test platform featuring magnet-relevant Lorentz forces and exhibits associated training. Here we report on the comparative behaviour of BOX samples comprising the same Nb$_3$Sn Rutherford cable but impregnated either with common resins used in high-field magnets, or with less conventional paraffin wax. Remarkably, the two paraffin wax-impregnated BOX samples reached their critical current without training and are also resilient to thermal and mechanical cycling. These rather encouraging results strongly contrast to those obtained with resin impregnated samples, which show the characteristic extensive training and at best barely reach their critical current value.

## 1. Introduction

Avoiding extensive training and/or performance degradation has proven to be a difficult challenge for epoxy-resin-impregnated high-field Nb$_3$Sn accelerator magnets wound from Rutherford cable [1]–[5]. Several recent magnets with designs ranging from canted-cosine-theta (CCT) [6]–[9] to classical cosine-theta [10, 11], exhibited severe training that complicates their application on a large scale in future accelerators. Their windings built with Nb$_3$Sn Rutherford cables were vacuum impregnated with epoxy resins reinforced with fiberglass. The cable impregnation with resin is necessary to suppress conductor movement and most importantly to provide well-supported conditions for the Nb$_3$Sn strands in order to avoid severe critical current degradation due to transverse stress working on the Nb$_3$Sn filaments [4, 5, 12].

Earlier research has highlighted the possibility of using alternatives to epoxy resins. Either bees- or paraffin wax, have been proposed and assessed on dipole, quadrupole and solenoid [13, 14] magnets with promising results. These Nb-Ti [15]–[18], Bi-2212 [18, 19] and $R$eBCO [20, 21] based magnets have shown in some cases substantial reduction in training, more reliable operation and reached high currents close to the critical current.

In this paper, the results for resin and wax impregnated BOnding eXperiment (BOX) samples are shown. The resins used are the so-called Mix 61 [23], CTD-101K [16], CTD-701X [16] and MY 750 [24], and the wax impregnated samples used paraffin wax. The epoxy resins selected for the comparison with wax have been used in the fabrication of high-field magnets and have varying elastic, plastic and fracture toughness properties [25]–[27] that may influence the training and overall performance of such magnets. Here we assess training, fatigue behaviour and resilience of paraffin wax over multiple powering cycles and a thermal cycle.

Previous experiments with BOX samples have shown that these can reproduce the characteristic training behaviour of accelerator magnets impregnated with a variety of epoxy resins in a cost-effective way [28]. The BOX experiments aim to seek the best impregnation materials and best practices primarily for CCT magnets, as well as assessing their suitability for other accelerator magnet designs.

**BOX sample fabrication**

The effectiveness of the impregnation system is assessed with a Rutherford cable made of 21 strands of B-OST RRP 108/127 $Nb_3Sn$ 0.85 mm in diameter [29] and insulated with fibreglass braid. A schematic of the sample is shown in Figure 1, and a detailed description of the BOX experiments can be found elsewhere [28].

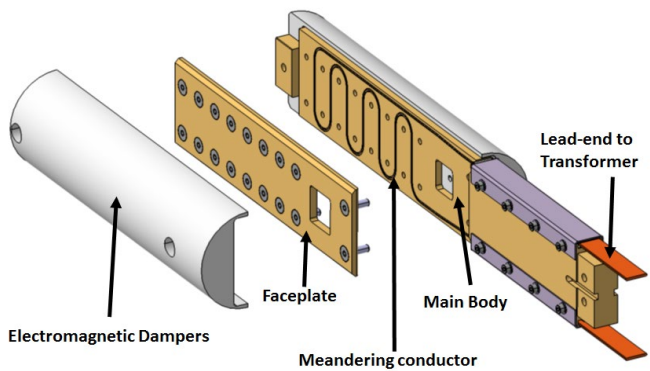

**Figure 1: Overview of the BOnding eXperiment (BOX) sample assembly, showing the meandering cable in the former, face plate and aluminium half shell dampers. Further details can be found in Daly et al. [28].**

All samples had the channel in which the cable is enclosed sand blasted. To remove most of the sizing on the fibreglass braid, cable and braid were washed for 10 minutes prior to winding using an ultrasonic bath of demineralised water with 1% Alconox detergent. Washing off the sizing helps to reduce carbon deposits in the channel during reaction heat treatment and improves bonding when impregnated with resin [28]. All samples went through the recommended heat treatment for $Nb_3Sn$ of 210°C for 72 hours, 400°C for 48 hours and 665°C for 50 hours in argon atmosphere. Prior to impregnation, the samples were dried under vacuum overnight. The impregnation of all samples occurred in a large vacuum vessel. The sample is contained within an "open top" container in the vacuum vessel, which is filled with resin or wax to ensure complete filling of cable and channel.

A commercially available paraffin wax POLARIT® G 54/56 from TH. C. TROMM GmbH, also used for candle-making, was chosen for the wax impregnation for its low melting point between 52-56°C and low viscosity ranging between 2-4 mPas at 100°C. The wax was degassed in a separate pot at 1 mbar, stirred and heated to 90°C. Prior to filling the sample, the mixing pot was brought to atmospheric pressure while maintaining the temperature between 80 to 90°C. The pressure in the impregnation vessel was kept above 3 mbar and the sample was kept at 80°C. The flow of wax was controlled to ensure slow filling thereby avoiding trapped gas as the wax fills the container and the sample from bottom to top. Once the container and sample were filled, they were slowly cooled down to room temperature over 6 hours. Two wax samples were produced following the same fabrication to verify reproducibility of the training test results. Once fully hardened, excess wax in the container around the sample was removed. Importantly, no noticeable recess of wax was detected within the channel of the samples when removing the face plate for visual inspection. The Wax-1 impregnated sample is shown in Figure 2a with the face plate removed. Both samples showed a good wax filling albeit with some local poor wetting of the fibre glass layer between the faceplate and the main body.

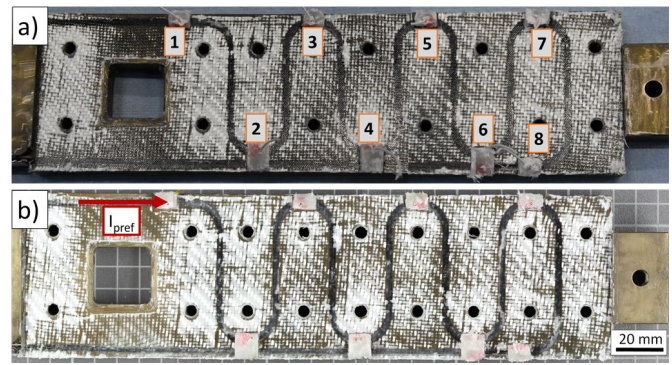

**Figure 2: Image of sample Wax-1 after impregnation (a), and after testing (b). The face plate is removed exposing the cable channel. Positions of voltage taps are shown in picture (a). Quenching segments are identified as the cable lengths between V-taps such as Segment 1-2, 2-3. 3-4, etc. The preferred current direction for the experiment is also indicated ($I_{pref}$).**

The resin impregnated samples were prepared using the same equipment and procedure but following their respective temperatures for processing and curing. No noticeable porosity or damage were noticed on the resin impregnated samples when removing the face plate after impregnation.

All samples were instrumented with 8 voltage taps to identify and locate the serpentine cable segments quenching the most.

## Test Results

The test program for the wax-filled samples consisted of verifying training behaviour up to the critical current and assessing thermal cycling and fatigue resilience in a background field of 7.5 T applied perpendicular to the wide surface of the straight cable segments yielding the highest Lorentz force on the cable at its critical current. All tests were performed at 4.2 K with a current ramp rate of 200 A/s. The preferred direction of the current ($I_{pref}$) for most tests was in the direction from V-taps 1 to 8. The training results for all samples are shown in Figure 3. The "fatigue and thermal cycle" behaviour for the wax filled samples are shown in Figure 4, and results of the current-reversal stability studies are shown in Figure 5 for the wax filled samples and three resin impregnated samples.

### 1. Training

Both wax filled samples did not exhibit any training and reached immediately a higher current of approximately 24.5 kA (Wax-1) and 24.6 kA (Wax-2) than the critical current of 23.8 ± 0.2 kA as determined by V-I measurements on a few segments. In sample Wax-1, segment 4-5 was primarily the quenching segment while in sample Wax-2, segment 3-4 was exclusively the quenching segment when powering the sample in the preferred current direction. Both segments are within the high and homogeneous magnetic field region of the solenoid.

All resin-impregnated samples exhibited extensive training as usually seen in resin-impregnated CCT magnets, and did not reach the high quench currents observed in the wax filled samples. The CTD-701X sample quenched first at a relatively high current of 18.1 kA while the Mix 61 sample quenched first at 8.9 kA. In addition, the resin impregnated samples clearly suffer from instabilities during the training with sometimes significant drops in current as known to happen in magnets with this type of $Nb_3Sn$ cable as well. To ensure reproducibility in training behaviour of the samples in this experimental setup, two samples impregnated with CTD-101K were fabricated and showed practically the same training behaviour.

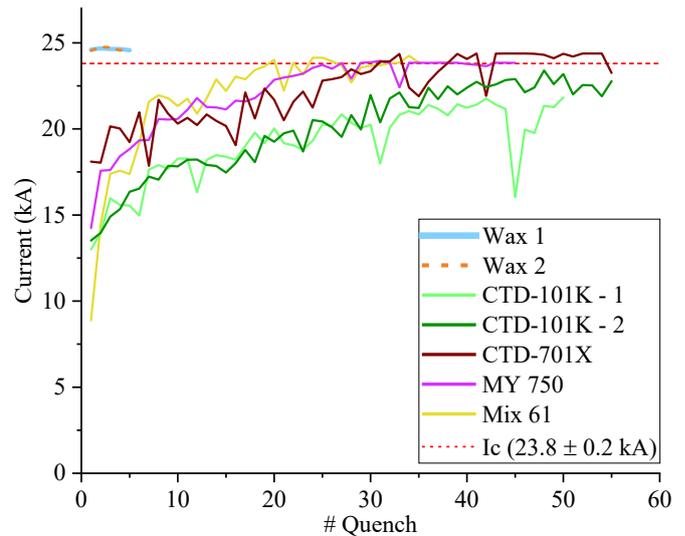

**Figure 3:** Training curves of 7 BOX samples. The two Wax filled samples (top left) show no training and notably reach values beyond the critical current at the first quench.

### 2. Fatigue behaviour by electro-mechanical (EM) cycling

The wax-filled samples were tested further for fatigue due to mechanical stress caused by Lorentz force by cycling the current between zero and 23 kA, a value just below the critical current, thereby avoiding a quench in the sample. The fatigue cycle runs typically lasted for 10, 30 and 50 consecutive ramping cycles. Once multiple cycles were completed, the samples were ramped up to quench again to check for degradation as shown in Figure 4.

The wax filled samples showed very little degradation over time and only dropped by 1 to 2% in maximum current over 60 to approximately 200 cycles. Some instability in segment 1-2 of sample Wax-1 was observed in the latter stages of cycling but it recovered. This first segment (1-2) behind the sample's electrical connection has also proven unstable in resin impregnated samples, which may indicate a sample entry effect independent of the impregnating medium used. This segment is closest to the un-impregnated leads located at the top end of the solenoid and within the more non-uniform solenoidal field region.

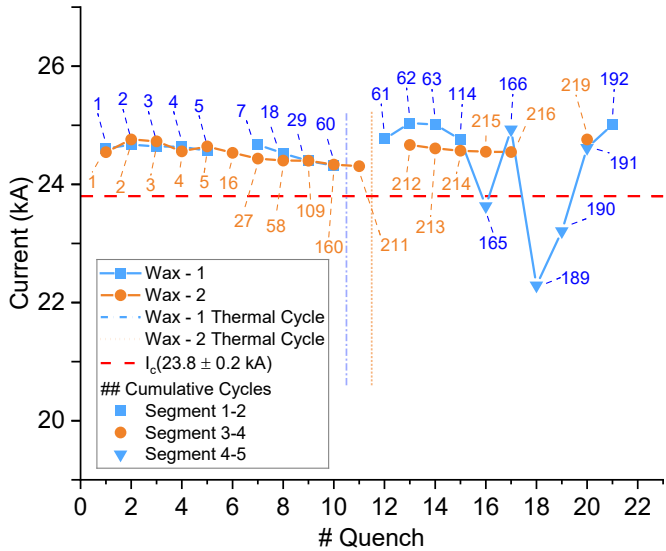

**Figure 4:** Quench current versus quench number of the two Wax filled samples with intermezzo's of multiple cycles up to 23 kA. The symbol markers indicate the quenches while the numerical markers show the number of intermediate energizing cycles at or above 23 kA.

*3. Thermal cycle behaviour*

Both wax samples were warmed up to room temperature after a number of electro-mechanical cycles, cooled back to 4.2 K and ramped again to maximum current. Sample Wax–1 underwent the single thermal cycle after 60 electro-mechanical cycles and sample Wax-2 after 211 cycles. Importantly, in both samples no retraining was observed and the maximum current reached was higher than earlier achieved in the last phase of testing prior to warm-up indicating a good memory and an absence of permanent degradation.

*4. V-I measurements for testing degradation*

V-I measurements for the determination of the critical current of the cables were performed on both wax-filled samples before cycling, after cycling and after thermal cycling as summarised in Figure 6. The wax filled samples showed equally good stability as the MY 750 resin impregnated sample with barely noticeable changes (within 1% error bars) in critical current indicating negligible conductor damage after EM cycling and a thermal cycle.

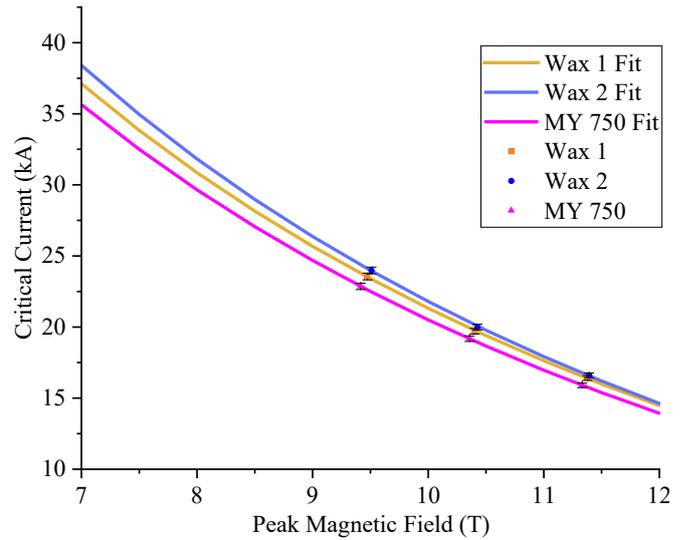

**Figure 5:** Critical current versus peak magnetic field of samples Wax-1, Wax-2, and MY 750. Dots are measured data with 1% error bars indicated, lines are a curve fit using the Godeke scaling law [30].

*5. Current reversibility studies*

The cable segments in the BOX layout exposed to Lorentz force are either pushed in or out of the channel depending on the direction of the current in the sample. This behaviour can be altered by changing the current direction. The resilience in training behaviour against current reversal was tested for both wax filled samples and compared to what was found with three resin impregnated samples. The results are shown in Figure 7. Sample Wax-1 immediately reached an equivalently high reverse quench current while sample Wax-2 had one quench in entry segment 1-2 before reaching a similarly high current. In clear contrast, the resin impregnated samples started to retrain in the opposite current direction, and in some cases retrained when the sample was powered in the original current direction. While force-reversal is not typically observed in accelerator magnets (with the exception of combined-function magnets), the experiment demonstrates the high degree of resilience of $Nb_3Sn$ Rutherford cable impregnated with paraffin wax.

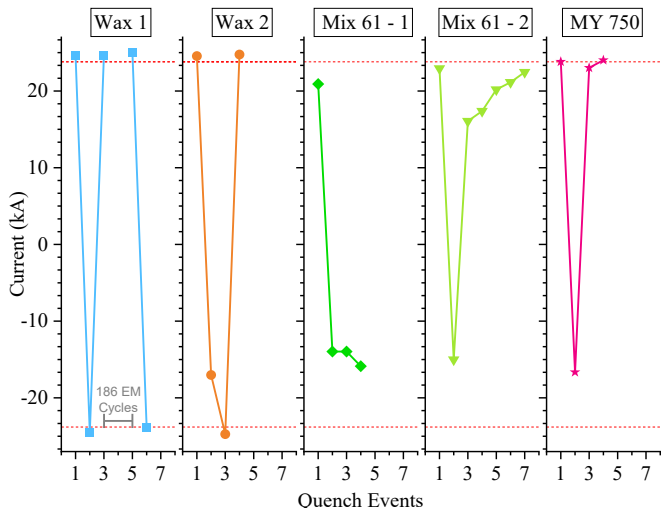

**Figure 6:** Result of current reversibility studies on the two wax filled samples, the MY 750 impregnated sample and two earlier Mix 61 impregnated samples [28].

As a final remark after testing the wax-filled samples we note that no clear damage was apparent on the surfaces of the samples as shown in Figure 2 a) and b). Subjectively, a slightly whiter "crazed" wax may be noticed on the straight segments 1-2, 3-4, 5-6 and 7-8 that were pushed out most frequently during the test with some small white patches possibly indicating minor detachment from the fibreglass interlayer.

### Discussion

At 7.5 T applied magnetic field and a cable current of 25 kA, the Lorentz force pushing the cable in or alternatively out of the channel in the sample holder is 6.6 kN, equivalent to 186 kN/m, and working on the 35.5 mm long central straight segments in the serpentine-shaped sample. This results in 11 MPa of mean shear stress at the wax/channel-wall interface assuming a fully bonded interface. Similarly, the pressure on the thin edge of a flat rectangular cable may be in the range of 30 to 135 MPa when assuming a fully bonded or a frictionless sliding interface, respectively.

The absence of training in the wax impregnated $Nb_3Sn$ cable samples is highly significant, but not necessarily unexpected as previous experiments using wax as impregnant for cables and coils showed promising results [17]. Importantly, these earlier experiments were performed on other conductors than $Nb_3Sn$.

Possible reasons for the absence of training and critical current degradation in the wax impregnated samples rely on a number of assumptions, namely:
- Low energy failure events as a result of the low fracture toughness of the wax showing typically a $K_{Ic}$ value that is a factor 10 lower at cryogenic conditions than in epoxy resins so far used in high-field magnets [31]–[34].
- Low shear modulus and minimal bonding between the wax and the channel wall and/or cable surface;
- Limited transmission of the failure energy within the insulating medium;
- The wax being fully confined within the channel and not allowed to flow out, thus providing sufficient mechanical support and mechanical stability for the conductor.
- Wax acts as a low-friction moderating medium, impeding any sudden conductor movement.

These assumptions are to be independently verified at 4.2 K in dedicated experiments to confirm values found in literature and to account for the presence of fibreglass in the composite.

In the next program step, the presented experiments will be complemented by testing the response to transverse compressive pressure on identical wax filled $Nb_3Sn$ cables within the University of Twente's test facility by which transverse pressure can be applied up to 200 MPa onto the broad face of the cable at 4.2 K, representative for the transverse pressure load on cables in 11 to 16-T high-field accelerator magnets [35].

Furthermore, the fatigue properties of the wax-filled cable samples deserve further investigation. Comparative studies with commonly used epoxy resins are to be performed to see if the measured behaviour is unique to waxes. Moreover, we plan to test new mixtures of epoxy resins in an attempt to reproduce wax-like training properties.

So far little literature information is present concerning the mechanical, radiation-resistance, and dielectric properties of paraffin wax and wax-glass composites at 4.2 K. These need to be investigated as well for feeding finite element models in order to enhance our understanding of the observed behaviour that allows wax impregnated samples to reach very high quench currents while eliminating training entirely for the BOX sample configuration.

### Conclusion

The impregnation with paraffin wax of a characteristic $Nb_3Sn$ Rutherford cable insulated with standard S2 fibreglass braid in BOX samples greatly contributed to eliminating training at up to 11 MPa mean shear stress on the cable broad face and up to 135 MPa of pressure on the short cable edge. Furthermore, the samples showed very good fatigue behaviour up to 200 electro-mechanical cycles and an excellent memory after one full thermal cycle to room temperature.

These new and remarkable results are in stark comparison with those of resin impregnated samples suffering from classical prolonged training and in some cases severe limitations in current.

The rather positive results presented in this paper highlight the possibility of practically eliminating training in certain $Nb_3Sn$ Rutherford cable based magnets and reaching critical current using paraffin wax. The V-I measurements on the wax samples highlighted further the stability offered by the fibreglass wax composite.

Further quantifiable investigations into the mechanical and fracture properties of paraffin wax are still required. The mechanisms of low energy failure and the fully confined cable/channel configuration that may drive the successful outcome of these wax impregnated BOX samples are yet to be fully understood.

## Acknowledgements

The authors are thankful for the private communication with Martin Wilson regarding earlier experiments on wax filled superconducting magnets at the Rutherford Laboratories, UK.